\documentclass[12pt]{article}

\usepackage{epsfig,multicol,multirow,cite}
\usepackage{amsmath,subfigure,latexsym,amssymb}
\usepackage[figuresright]{rotating}
\usepackage{tikz, graphicx}
\usepackage[utf8x]{inputenc}

\usepackage[colorlinks=true,citecolor=magenta,urlcolor=blue]{hyperref}
\usepackage{hyperref}

\newcommand{\be}{\begin{equation}}
\newcommand{\ee}{\end{equation}}
\newcommand{\nn}{\nonumber}
\newcommand{\bea}{\begin{eqnarray}}
\newcommand{\eea}{\end{eqnarray}} 
\newcommand{\eps}{\epsilon}
\newcommand{\la}{\langle}
\newcommand{\ra}{\rangle}

\newcommand{\R}{{\kern+.25em\sf{R}\kern-.78em\sf{I} \kern+.78em\kern-.25em}}
\newcommand{\RR}{{\kern+.25em\sf{R}\kern-.6em\sf{I} \kern+.6em\kern-.25em}}
\newcommand{\N}{{\kern+.25em\sf{N}\kern-.78em\sf{I} \kern+.78em\kern-.25em}}

\usepackage{color}
\definecolor{col_red}{rgb}{1.0,0.0,0.0}

\makeatletter
\@addtoreset{equation}{section}
\makeatother

\begin{document}
 
\begin{center}
{\Large\bf Zurek's scaling-law in the Ising model}

\vspace*{6mm}

{\Large\bf out of thermal equilibrium} \\

\vspace*{1cm}

Jos\'{e} Armando P\'{e}rez-Loera and Wolfgang Bietenholz
\\
\ \\
Instituto de Ciencias Nucleares \vspace*{0.7mm} \\
Universidad Nacional Aut\'{o}noma de M\'{e}xico \vspace*{0.7mm} \\
A.P.\ 70-543, C.P.\ 04510 Ciudad de M\'{e}xico, Mexico\\

\end{center}

\vspace*{6mm}

\noindent
The Kibble mechanism plays a prominent role in the theory of the
early Universe, as an explanation of the possible formation of
cosmic strings. Zurek suggested the analogous effect in liquid
helium under rapid cooling, and he conjectured --- together with
del Campo and Kibble --- a scaling-law for
the relation between the density of remnant topological defects and
the cooling rate, when a system passes through its critical temperature.
Such scaling-laws were indeed observed in condensed matter experiments.

Here we test the validity of Zurek's scaling-law in a different
framework. We numerically study the behavior of the Ising model
(with classical spins) in 1, 2 and 3 dimensions under rapid cooling.
This model does not have topological defects, so we consider the
dynamics of domains instead, {\it i.e.}\ we measure their evolution
during a cooling process down to zero temperature.
For several Markov chain Monte Carlo algorithms, we consistently
observe scaling-laws along the lines of Zurek's conjecture, in all
dimensions under consideration, which shows that this feature holds more
generally than expected. It is highly remarkable that even the exponents of
these scaling-laws are consistent for three different algorithms,
which hints at a physical meaning.

\newpage

\section{Motivation}

The setup of the Ising model is very simple, and yet it is
one of the most popular models in Statistical Mechanics. It
has been extensively studied in thermal equilibrium, with
well-known analytic solutions of the 1d (one dimensional) \cite{Ising}
and 2d model \cite{Onsager}, and high precision results based
on numerical simulations in $d=3$ dimensions, in particular for
the critical temperature and critical exponents, see {\it e.g.}\
Ref.\ \cite {3dIsingsimu} and references therein.\footnote{Of course,
there are also numerical studies and precise results in higher
dimensions, but our work refers to $d \leq 3$. Moreover, in
$d \geq 3$ there are also successful analytic approximation methods.}
It can also be analyzed by mean field theory, which corresponds to
the limit of an infinite number of dimensions.

On the other hand, little is known about the Ising model
out of thermal equilibrium, which seems --- at least in
$d > 1$ --- hardly tractable by analytical approaches.
Here we present simulation results for this model under rapid
cooling in 1, 2 and 3 dimensions, obtained with three algorithms.

The physical motivation for this study refers to the Kibble-Zurek
mechanism. Kibble formulated this concept in the context of the
early Universe \cite{Kibble}: during the electroweak phase transition,
about $10^{-12}$~sec after the Big Bang,
the Higgs field acquired a non-zero vacuum expectation value, which came
along with arbitrary complex phases. In regions of the Universe, which
were causally disconnected, these phases could not be correlated. In the
interfaces between such regions, topological defects are naturally
expected. Under adiabatic cooling they would disappear,
but under the very rapid cooling in the early Universe, part of
them may well have survived. Kibble further conjectured that
such topological defects might have persisted to this day,
due to their topological stability, and lines
of such defects could still penetrate the Universe as cosmic
strings, as reviewed {\it e.g.}\ in Ref.\ \cite{KibbleReview}.

In the past century, the observational search for cosmic strings
focused on anisotropies in the Cosmic Microwave Background, which
led to bounds for their string tension $\mu$ of the magnitude
$G \mu < {\cal O}(10^{-7})$ \cite{CMB} (where $G$ is the gravitational
constant). This rules out their role as seeds for galaxy formation,
which was the original hypothesis, but of course it does not rule
out the existence of cosmic strings.

The successful detection of gravitational waves provides
new perspectives for the search for evidence of cosmic strings.
Hence the subject is timely again, and the LIGO-Virgo-KAGRA
Collaboration reports a drastically improved bound of
$G \mu \lesssim 4 \times 10^{-15}$ for oscillating loops of cosmic
strings \cite{LIGO}. The space-based LISA Observatory hopes for
a future sensitivity up to $G \mu = {\cal O} (10^{-17})$ \cite{LISA}.

The observation of cosmic strings is not successful so far, but the
corresponding mechanism has been observed in the laboratory, as
suggested by Zurek \cite{Zurek}.
In particular, vortices in superfluid helium ($^{4}$He) represent
topological defects of a similar kind. Under adiabatic cooling
down to very low temperatures they should disappear, whereas part
of are expected to them persist under rapid cooling.
Zurek conjectured a scaling-law,
according to which the remnant vortex density is proportional
to the cooling rate to some power. This was recently observed in a
Fermi gas undergoing a transition from normal fluid to superfluid
\cite{Lee24}, which is in the universality class of liquid helium.
Del Campo, Kibble and Zurek further conjectured the value of the
corresponding exponent at a specific final temperature \cite{DelCampo1314},
which is also reported to be
in agreement with the Fermi gas experiment \cite{Lee24}.

Further experimental evidence was observed in systems including
the defect density in a non-linear optical system \cite{Ducci99},
superconductors \cite{Monaco} (referring to the fluxon rate in
annular Josephson junctions after a rapid transition to
superconductivity), and several hexagonal manganites \cite{Griffin12,Lin14}.

There have been numerical simulations which confirm the existence
of such a scaling-law for the vortex density in the 2d \cite{JelCug}
and 3d \cite{Lin14,3dO2} XY model (or O(2) non-linear $\sigma$-model).

Those settings refer to topological defects --- which are dense above
a critical temperature $T_{\rm c}$ --- and their partial persistence when
the temperature quickly decreases below $T_{\rm c}$. The situation is
different in the Ising model, which does not have topological
defects, and in the 1d case it does not even undergo a phase transition.
However, the walls between uniform domains bear some similarity
to such disordering defects.

Here we study the persistence of Ising domains and their walls by means
of Monte Carlo simulations. We are particularly interested in the
question whether they follow a scaling-law in the spirit of Zurek's
conjecture, although that referred to different situations. Hence we
explore the question whether or not such scaling-laws could be valid
more generally than assumed.

Unlike equilibrium simulations, the results out of equilibrium generally
depend on the algorithm, so one might question how to relate them to
physical phenomena. Here we employ three different Monte Carlo algorithms
(with local updates or with cluster updates), since we are particularly
interested in properties, which coincide for different algorithms.

From the experimental side, a recent investigation of 3d Ising-type
domains in NiTiO$_{3}$ and BiTel
confirms scaling-laws for these domains and, in the first case,
also the predicted exponent \cite{Du23}. In addition, there are a
number of studies of the 1d quantum Ising model (quantum spin chain),
both from the theoretical (analytical and numerical) perspective
\cite{QIsing1d} and based on an experiment with an ion chain quantum
simulator \cite{Li23}. Also the classical 1d Ising model was
considered before, regarding the kink density in a spin chain
under Glauber dynamics \cite{Mayo21,Jindal24}.

Section 2 describes the setup, the observables and numerical methods,
Section 3 presents our simulation results, and Section 4 is devoted
to our conclusions.

\section{Model, observables and algorithms}

\subsection{Ising model}

We deal with the well-known Ising model, with classical spin
variables and with the Hamilton function
\be \label{Hami}
{\cal H}[s] = -J \sum_{\la xy \ra} s_x s_y \ , \quad s_x \in \{-1, 1\} \ ,
\ee
where $x,y$ are sites of a lattice with spacing 1 (lattice units)
with a square and cubic structure in $d= 2,3$, respectively, and a
ferromagnetic coupling $J>0$. The sum runs over nearest-neighbor
lattice sites, and $[s]$ denotes a spin configuration.
We use lattice volumes $L^d$ with periodic boundary
conditions, which provide (discrete) translation invariance.

The partition function is given by
\be 
Z(T) = \sum_{[s]} e^{-{\cal H}[s]/(k_{\rm B}T)} \ , 
\ee
where $T$ is the temperature and $k_{\rm B}$ the Boltzmann constant.
Since $Z$ only depends on $J/(k_{\rm B}T)$, we choose the temperature units
such that $J/k_{\rm B}=1$, for convenience (in the framework of lattice
units, there is no problem with the dimensions).

In infinite volume, the correlation length diverges at temperature
$T_{\rm c}$,
\bea
T_{\rm c} &=& 0 \qquad (d=1) \ , \nn \\
T_{\rm c} &=& \frac{2}{\ln (1 + \sqrt{2})} = 2.269185\dots \qquad (d=2) \ ,
\nn \\
T_{\rm c} &=& 4.511424 \dots \qquad (d=3) \ ,
\eea
which is the critical temperature in $d=2$ and $3$; below $T_{\rm c}$
the global $Z(2)$ symmetry is spontaneously broken.\footnote{In $d=1$,
  the expression ``critical temperature''
is not always used, since in that case there is no phase transition.}
The values of $T_{\rm c}$ were obtained analytically in $d=1$ \cite{Ising}
and $d=2$ \cite{Onsager}, and numerically in $d=3$, see {\it e.g.}\
Ref.\ \cite{3dIsingsimu}.

\subsection{Observables}

Standard observables are the energy ${\cal H}[s]$ and the
magnetization $M[s] = |\sum_{x} s_{x}|$, which we are going to consider.

We focus, however, on {\em domains} of parallel spins,
{\it i.e.}\ sets of spins with the same sign, which are (directly or
indirectly) connected by lattice links. Due to the periodic boundary
conditions, a domain can extend across one or several boundaries.
Links connecting opposite spins represent the domain
walls.\footnote{For the sake of an unambiguous terminology,
we refer to domain {\em walls}, but to {\em boundaries} of the
entire lattice volume.}
In a given configuration $[s]$, the total size of all domain walls,
$W[s]$, is related to the Hamilton function, or energy, as
\be
W[s] = {\cal H}[s] + d L^d \ .
\ee
In particular, we are going to consider the average domain wall size,
\be
\overline{DWS}[s] = W[s] / D[s] \ , \quad
D[s] = {\rm number~of~domains} .
\ee
Here the average refers to all the domains in one configuration.
This observable is somewhat unusual, but appropriate for the
purpose of this study, cf.\ Section 1. It represents an
independent observable in $d \geq 2$ (in $d=1$, it is just the
(even) number of kinks, which is equal to ${\cal D}[s]-1$).
These walls quantify the deviation from uniform order,
roughly in analogy to topological defects (such as vortices,
or monopoles) in other models.

\subsection{Algorithms}

For our Markov chain Monte Carlo simulations, we apply three different
algorithms. For the usual simulations in thermal equilibrium (at
fixed temperature $T$), this is not motivated, since they all lead
to the same results, and it is sufficient to use the most efficient
algorithm. For rapidly varying $T$, however, they are not equivalent
anymore; in this case, the results can depend on the algorithm, and it is
of major interest to search for properties, which are 
algorithm independent, thus permitting a physical interpretation.

We first consider two local update algorithms. Here one suggests the
sign flip of one spin, $s_x \to s_x' = -s_x$, which changes the
Hamilton function by $\Delta {\cal H} = {\cal H}[s']- {\cal H}[s]$.
In the (famous) Metropolis algorithm \cite{Metropolis} and the
(less known) Glauber algorithm \cite{Glauber}, this flip is accepted
with probability
\bea
p[s \to s']_{\rm Metropolis} &=& \left\{ \begin{array}{ccc}
  1 && {\rm if}~\Delta {\cal H} \leq 0 \\
  \exp (-\Delta {\cal H}/T) && {\rm if}~\Delta {\cal H} > 0
\end{array} \right. \ , \\
p[s \to s']_{\rm Glauber} &=& \frac{e^{-\Delta {\cal H}/T}}
{1+e^{-\Delta {\cal H}/T}} \ .
\eea
In equilibrium, both prescriptions provide detailed balance. Running
over the volume (in lexicographic order) and updating each spin in
this manner is denoted as one {\em sweep}.

We further apply the cluster algorithm, which is non-local, since
it updates entire clusters of spins. This algorithm first
considers pairs of nearest neighbor spins: if they are parallel,
we connect them by a {\em bond} with probability
$p_{\rm bond} = 1 - \exp(-2/T)$. A {\em cluster} is a set of spins,
which are connected (directly or indirectly) by bonds (it can
extend across boundaries). Hence a cluster can maximally coincide
with a domain (but it can also consist of just one spin).
Such a cluster can then be updated collectively, which means that
all its spins change their sign simultaneously.

The original multi-cluster variant, the Swendsen-Wang
algorithm \cite{SwenWang}, divides the entire volume into clusters,
and flips each cluster with probability $1/2$. This represents a powerful
sweep, which is actually too efficient for our purposes: even if the
temperature changes quite rapidly, its Markov chain evolution
is almost adiabatic, which does not allow us to study the analogue
of the Kibble-Zurek mechanism.

Hence it is favorable to use the single-cluster variant of this
algorithm \cite{Wolff}. Here one
chooses a random spin as a ``seed'', and builds just the one
cluster which includes this spin. This cluster is then flipped
for sure. This represents an update step, which is typically
much less powerful than a multi-cluster sweep (but faster to
perform). Hence it defines a shorter Markov time unit, which is
better suitable for monitoring a cooling process. We also denote
a single-cluster flip as one sweep; its quantitative comparison
to a local update sweep is not obvious.

\subsection{Simulations out of thermal equilibrium}

A simulation begins with a comprehensive thermalization at the
initial temperature $T_{\rm i}$.  This could be done with any of
these algorithm (they all fulfil detailed balance and ergodicity),
so here we use a cluster algorithm, since it is most efficient. 

The idea of the Kibble-Zurek mechanism is to pass from well above to
well below a critical temperature. In $d=2$ and $3$, this is achieved
by choosing $T_{\rm i} = 2 T_{\rm c}$ and the final temperature $T_{\rm f}=0$.
In $d=1$ we can only begin at some high temperature, which we choose as
$T_{\rm i} = 6$, and we reduce it again to $T_{\rm f} = 0$.

Once we have a well thermalized configuration, we begin the cooling
process down to temperature $T_{\rm f} = 0$, lowering $T$ after each sweep.
As in Refs.\ \cite{Lin14,JelCug,3dO2},
the temperature is reduced linearly\footnote{The assumption of a
{\em linear} cooling is common in the literature, {\it e.g.}\
in Refs.\ \cite{Zurek,DelCampo1314}.}
over a Markov time period $2 \tau_{Q}$,
in units of sweeps of the corresponding algorithm. We denote the parameter
$\tau_{Q}$ as the {\em inverse cooling rate} (in the
literature, it is also known as the ``quench rate'').
Thus the temperature dependence on the Markov time $t$ takes the form
\be  \label{Ttlin}
T(t) = \frac{1}{2} T_{\rm i}
\Big( 1 - \frac{t}{\tau_{Q}} \Big) \ , \quad t = [-\tau_{Q}, \tau_{Q}] \ .
\ee
In $d = 2$ and $3$, the coefficient $\frac{1}{2} T_{\rm i} = T(0)$ is the
critical temperature $T_{\rm c}$.

This cooling process\footnote{The term ``cooling'' is used here in
  a sense, which differs from the lattice literature.
  The more literal meaning describe above is sometimes also
  called ``quenching'',
  but --- for instance in lattice QCD --- that term has a different
  meaning as well.}
is repeated a large number of times, starting
each time from a different configuration, which is well thermalized
at $T_{\rm i}$. We take expectation values as mean values of the
configurations obtained at the same Markov time, {\it i.e.}\
at a fixed number of sweeps (after thermalization).

Since all our cooling processes end at $T_{\rm f} = 0$, an adiabatic
cooling (for very large $\tau_{Q}$) leads to a uniform configuration,
with energy density $\eps = {\cal H}/V = -d$, saturated magnetization
density $m = M/V = 1$,
and just one domain without wall. However, the faster the cooling,
the more do the expectation values --- at the final Markov time
$t= \tau_{Q}$ --- deviate from that trivial case.
The simulation results, to be presented in the next section,
provide quantitative data for this effect, which we discuss
in light of the Kibble-Zurek mechanism.

\section{Numerical results}

Table \ref{tabpara} displays the lattice volumes and the statistics
of numerical cooling experiments in each dimension. Further numerical
experiments in different volumes show that the finite-size
effects of our results are negligible. The statistical errors, which
are indicated in the following figures, were calculated with the
jackknife method.
\begin{table}[h!]
\begin{center}  
\begin{tabular}{|c|c|c|c|}
\hline
dimension $d$ & volume $V=L^{d}$ & cooling experiments & $T_{\rm i}$ \\
\hline
1 & 10,000   & 10,000 & 6 \\
2 & $150^{2}$ & 5,000  & 4.538370 \\
3 & $50^{3}$  & 1,000  & 9.022848 \\
\hline
\end{tabular}
\end{center}
\vspace*{-2mm}
\caption{Overview of our lattice volumes $V$ and the statistics, {\it i.e.}\
the number of the numerical cooling experiments --- with each algorithm ---
in different dimensions $d$, as well as the initial temperature
$T_{\rm i}$ (in lattice units and with 
$J/k_{\rm B} = 1$). In all cases, the final temperature is
$T_{\rm f} = 0$.\label{tabpara}}
\end{table}

\begin{figure}[h!]
\begin{center}
\includegraphics[scale=0.3]{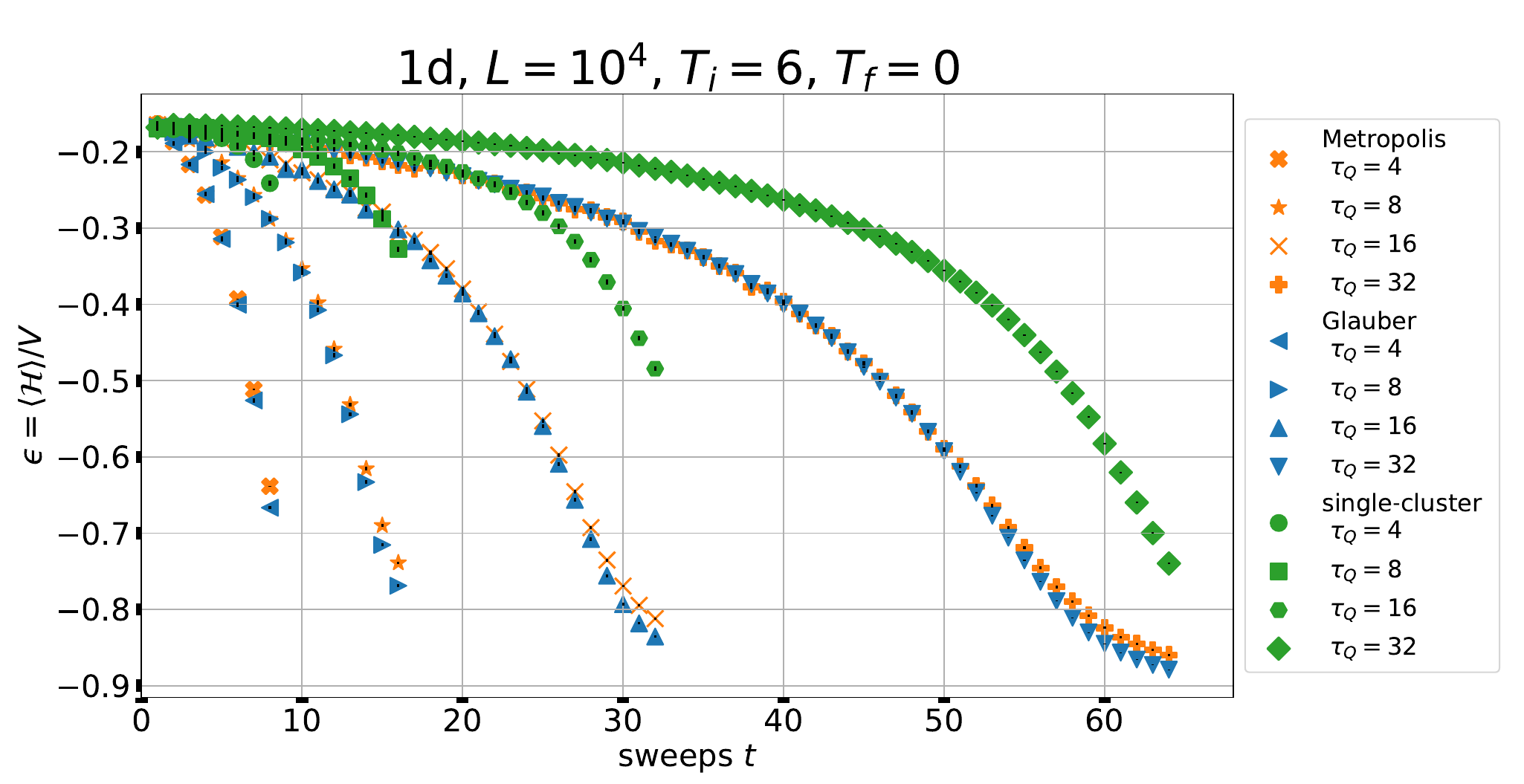}
\includegraphics[scale=0.3]{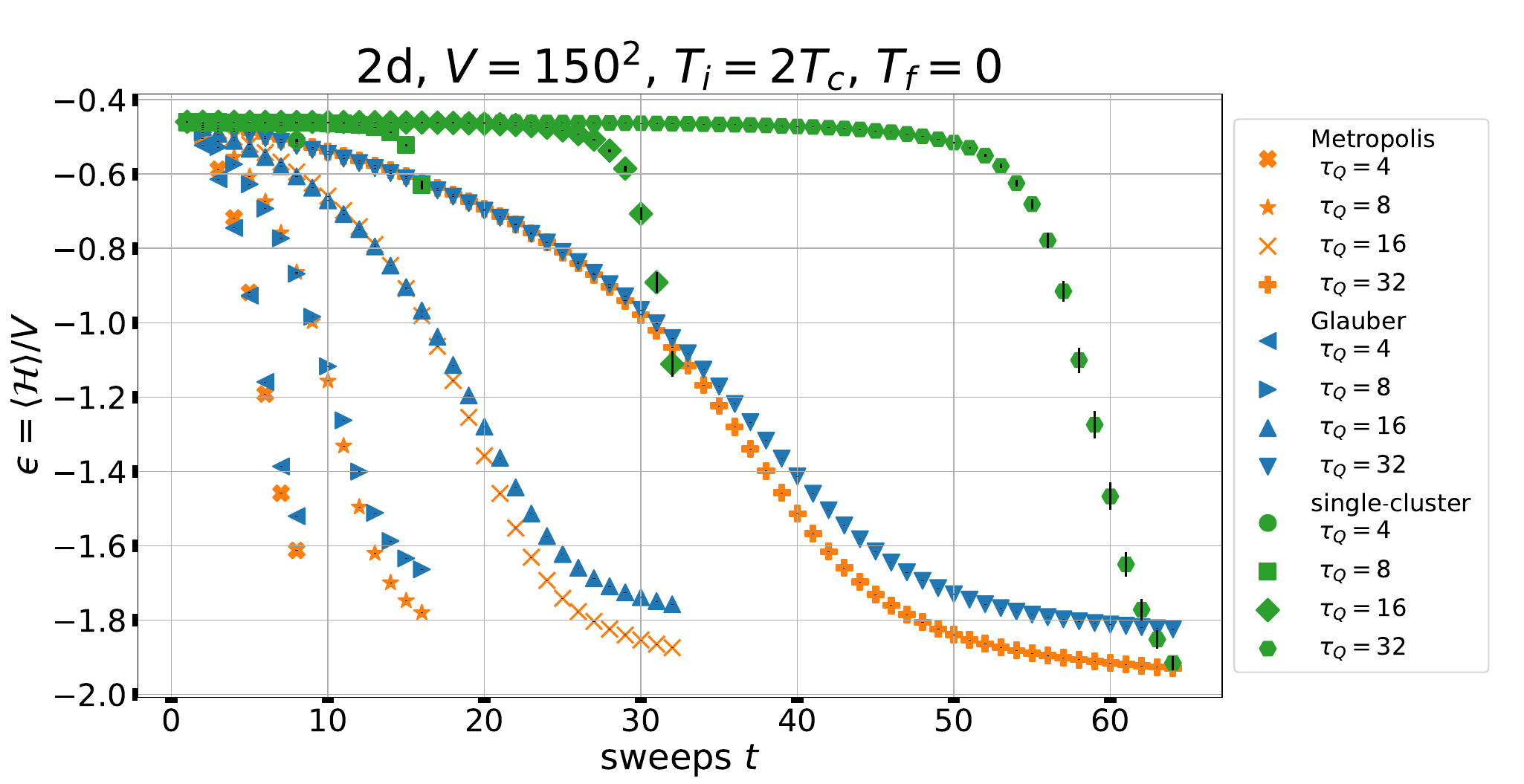}
\includegraphics[scale=0.3]{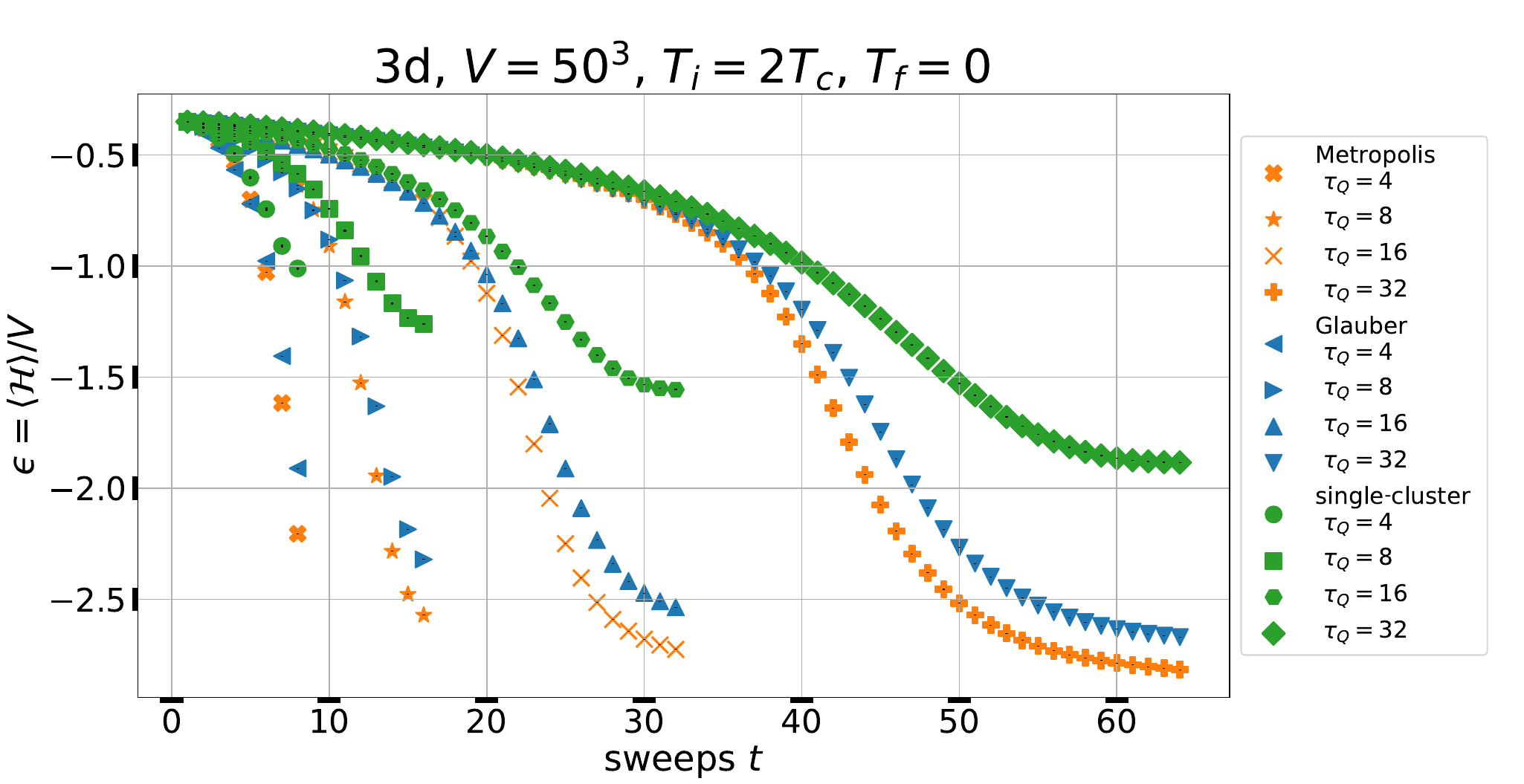}
\caption{\small The Markov time evolution of the energy density
$\epsilon = \la H \ra/V$ during the cooling processes, according to
eq.\ (\ref{Ttlin}), in $d=1$ (top), $d=2$ (center) and $d=3$ (bottom).}
\label{Edense}
\end{center}
\vspace*{-4mm}
\end{figure}

Figure \ref{Edense} shows the energy density $\epsilon = \la {\cal H}\ra/V$
as a function of the Markov time during the cooling process,
$\epsilon (t)$, $t= -\tau_{Q}, \dots ,\tau_{Q}$, in dimensions
$d=1,2,3$.\footnote{The expectation values denoted by $\la \dots \ra$
refer to a large set of independent configurations, which are
stochastically generated under identical conditions.}
The slowest cooling in this study ($\tau_{Q} =32$) is closest to
the adiabatic process,
hence it yields $\epsilon$-values below the faster cooling values
at the same temperature. Even at $T=0$, however, the final energy
density is still well above its minimum of $\epsilon = -d$.

The local update algorithms, Metropolis and Glauber, lead to similar
behaviors. In $d>1$, $\epsilon$ tends to be somewhat lower for the
Metropolis algorithm, which shows that it adjusts more rapidly to a
temperature change. This is consistent with the fact that it is more
efficient in equilibrium simulations.

For the single-cluster algorithm, the Markov time unit of one sweep
is difficult to compare to the local update algorithms (as we
mentioned before), but one sweep updates less spins in this case.
However, the exponents in the scaling-laws --- that we are ultimately
interested in --- do not depend on the Markov time units.
If we just compare the data in Figure \ref{Edense} for the
single-cluster algorithm and various inverse cooling rates $\tau_Q$,
the qualitative behavior is confirmed: faster cooling implies a
significantly higher energy density $\epsilon$ at a fixed
temperature $T < T_{\rm i}$.

\begin{figure}[h!]
\begin{center}
\includegraphics[scale=0.3]{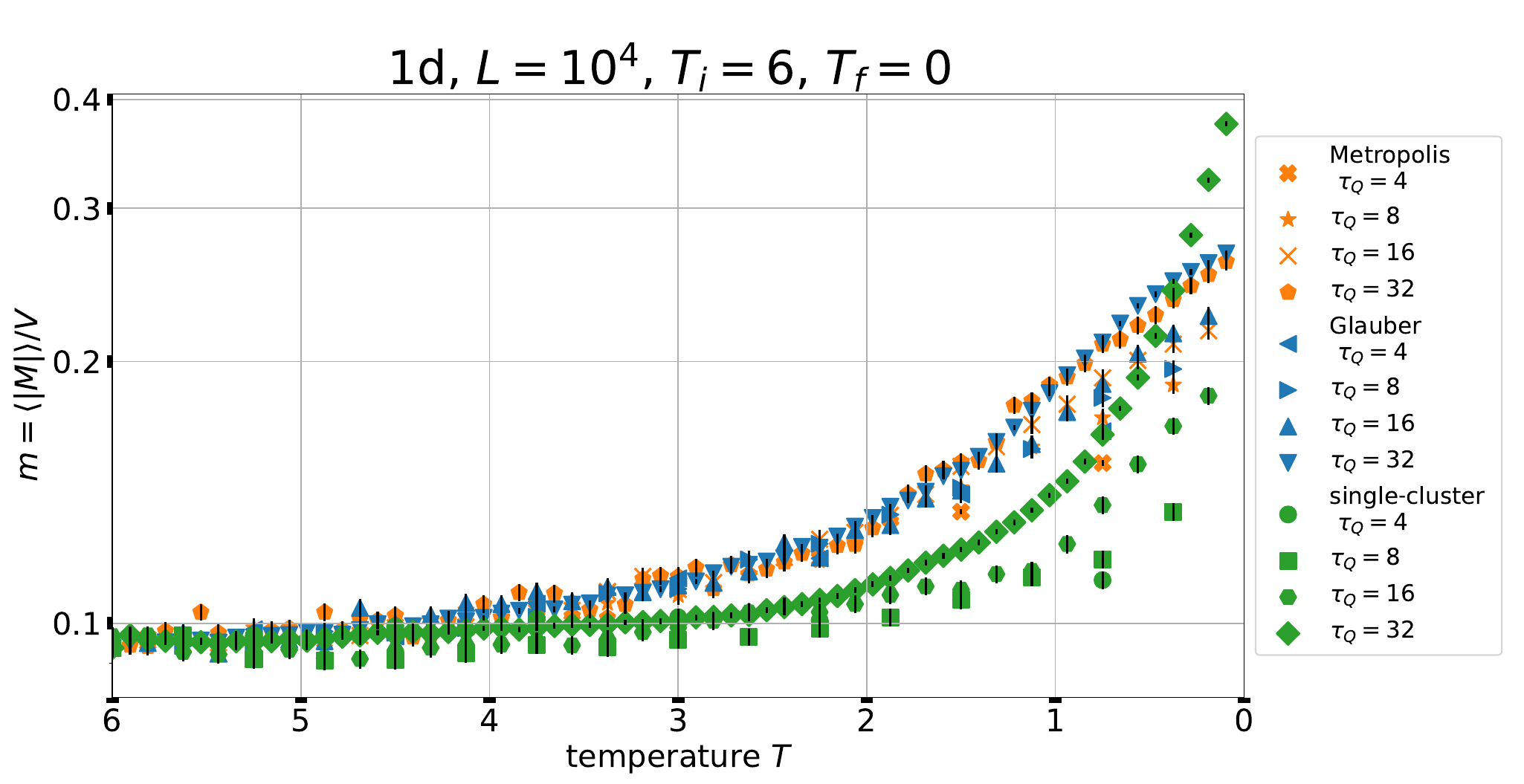}
\includegraphics[scale=0.3]{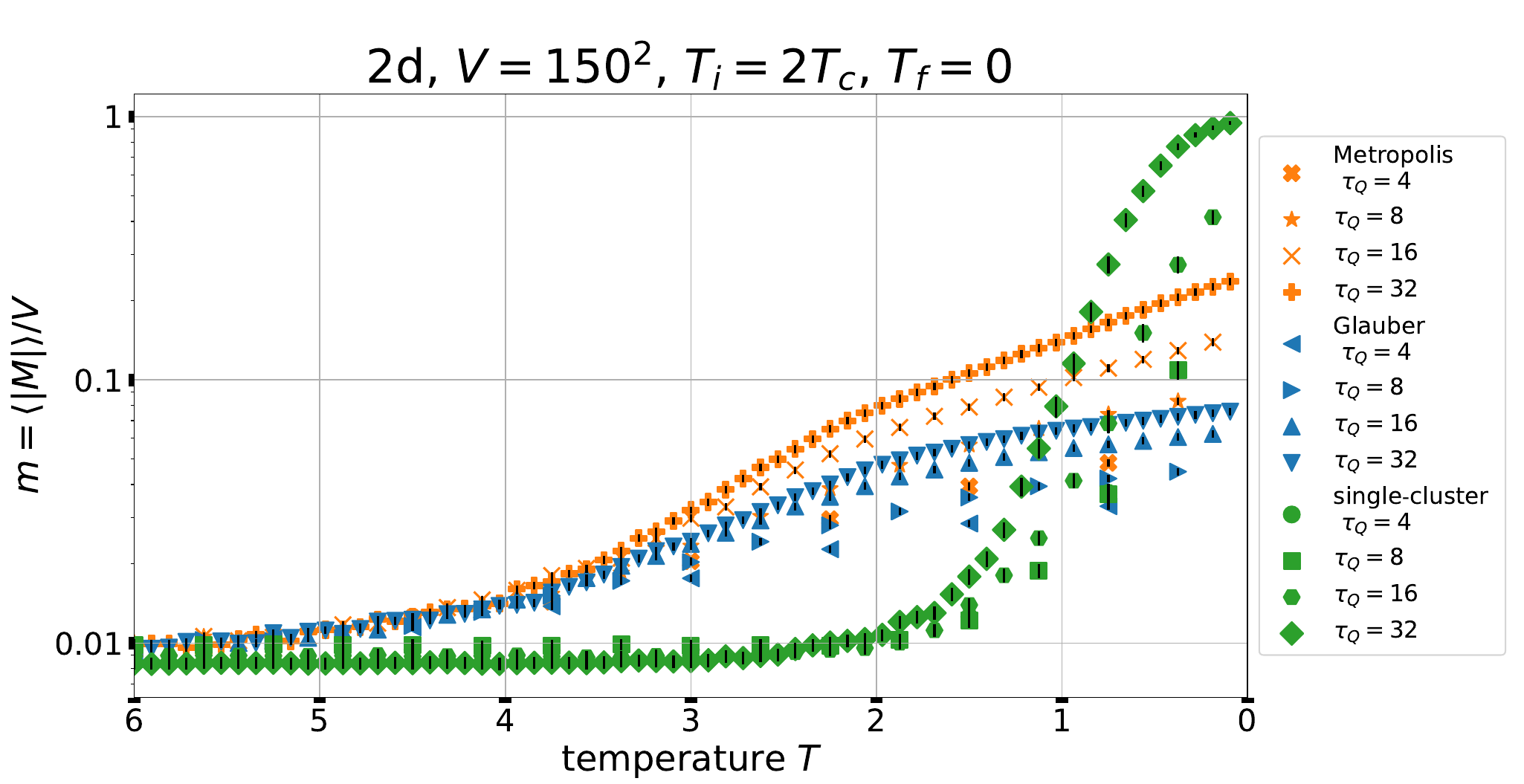}
\includegraphics[scale=0.3]{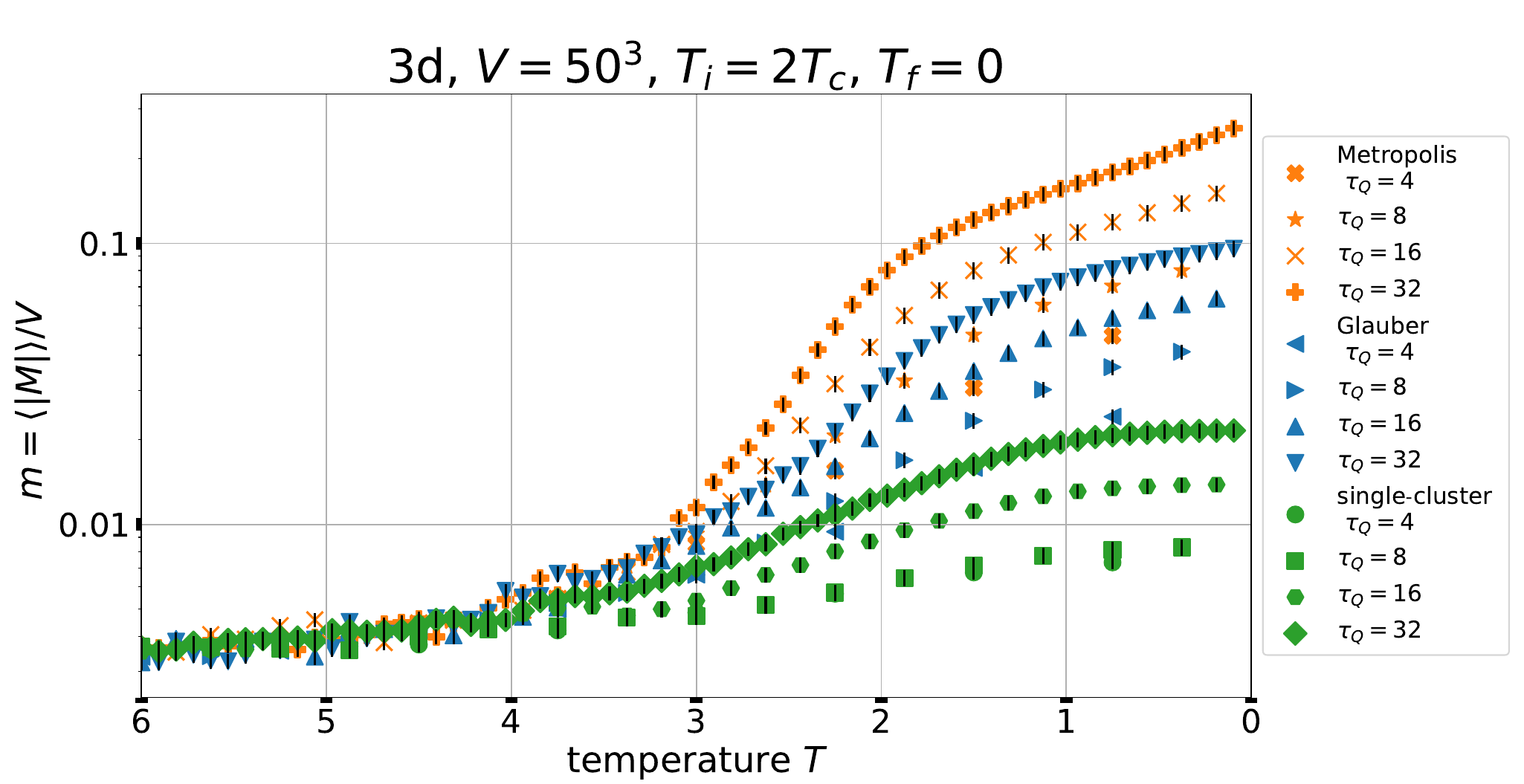}
\caption{\small The Markov time evolution of the magnetization density
$m = \la M \ra/V$ during the cooling processes, as a function of the
temperature $T$, in $d=1$ (top), $d=2$ (center) and $d=3$ (bottom).}
\label{Mdense}
\end{center}
\vspace*{-4mm}
\end{figure}

Figure \ref{Mdense} shows the Markov time evolutions of the magnetization
density $m = \la M[s] \ra /V = \la |\sum_{x} s_{x} | \ra /V$. This
quantity is similar to the energy, but it increases, starting from
a value close to 0 (at high $T$) up to 1, minus the remnant disorder
due to the fast cooling. We show $m$ as a function of $T$, which
illustrates that --- at the same temperature --- the values are less
$\tau_{Q}$-dependent than it might seem in Figure \ref{Edense},
although the qualitative trend of this dependence is the same.

\begin{figure}[h!]
\begin{center}
\includegraphics[scale=0.3]{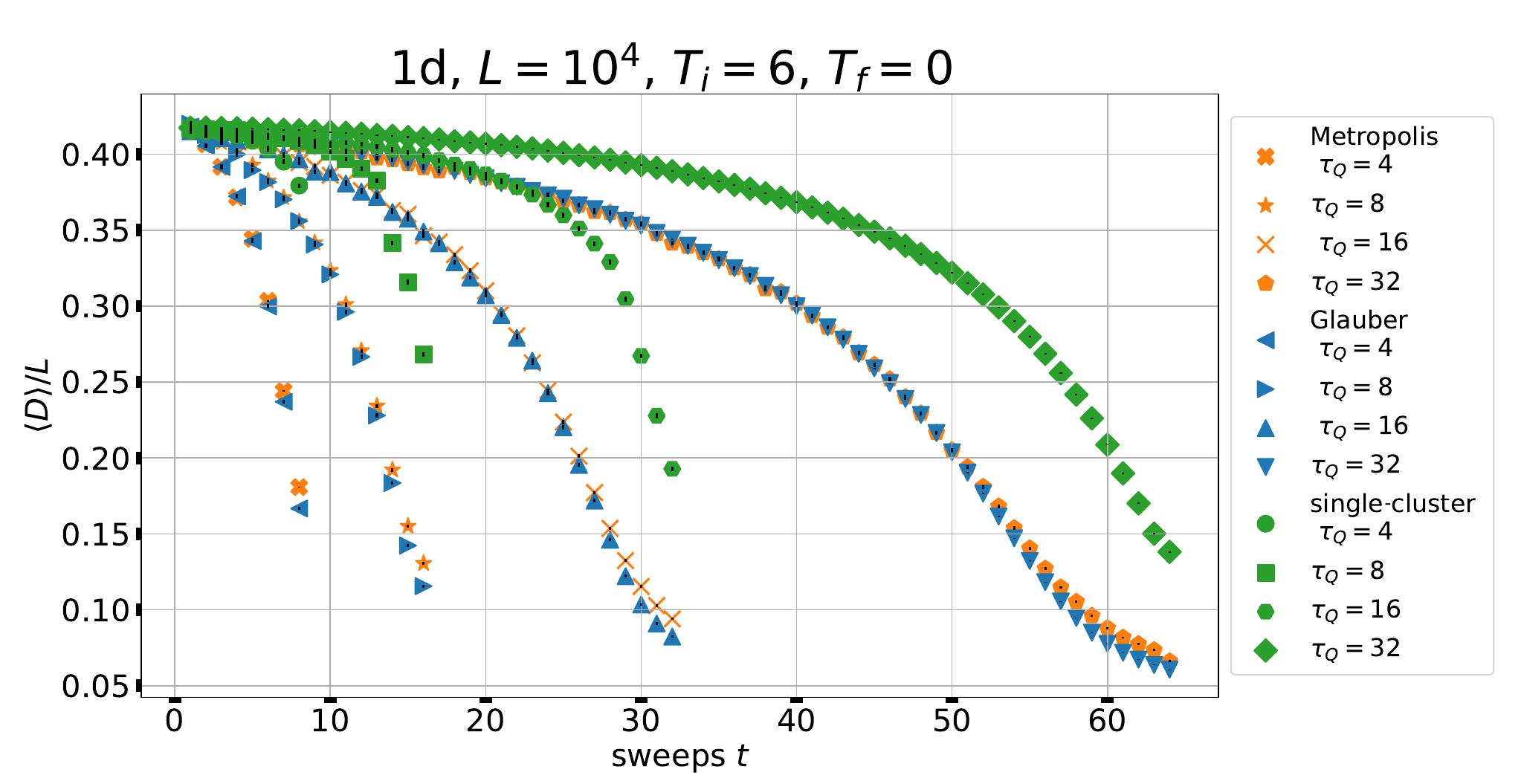}
\includegraphics[scale=0.3]{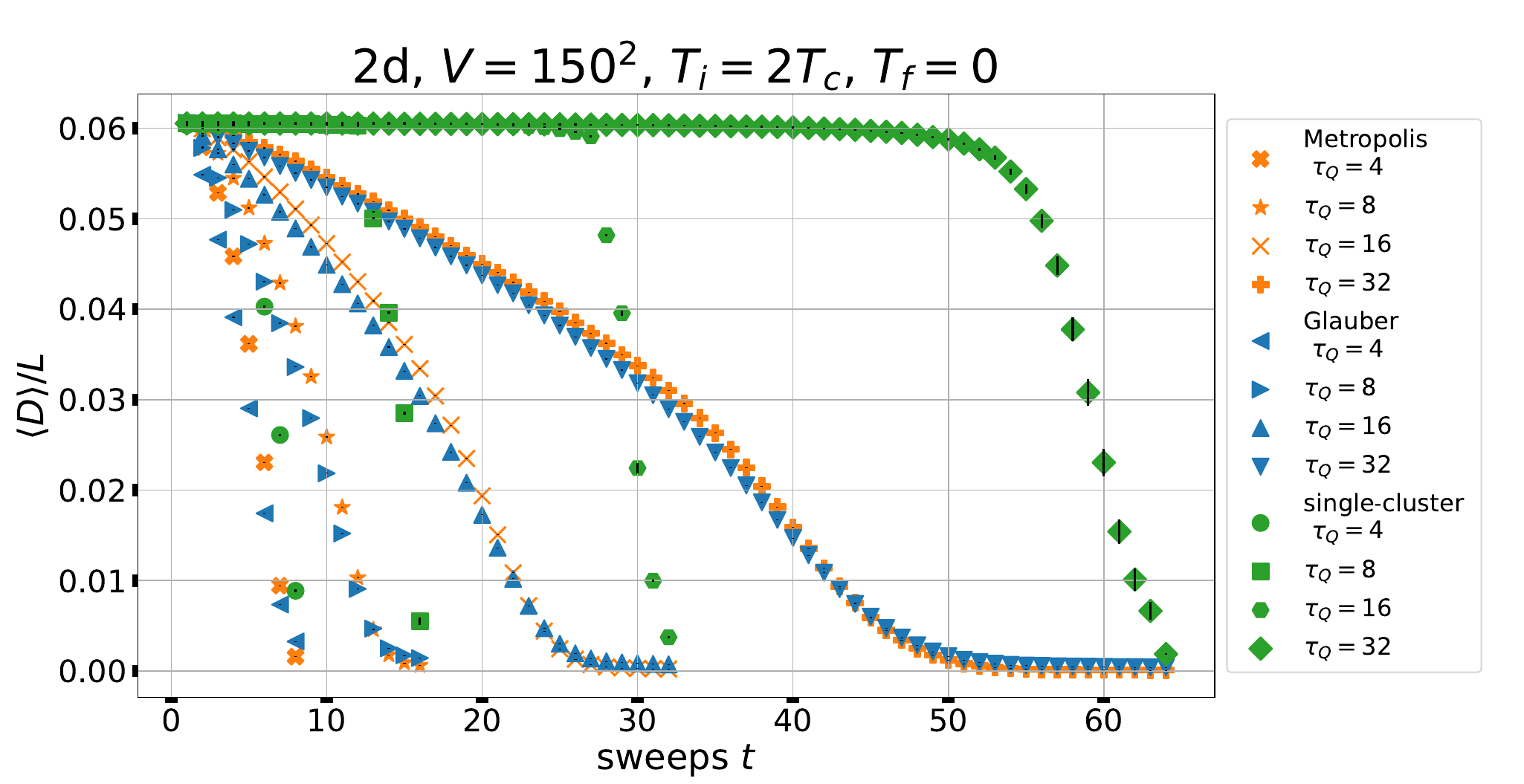}
\includegraphics[scale=0.3]{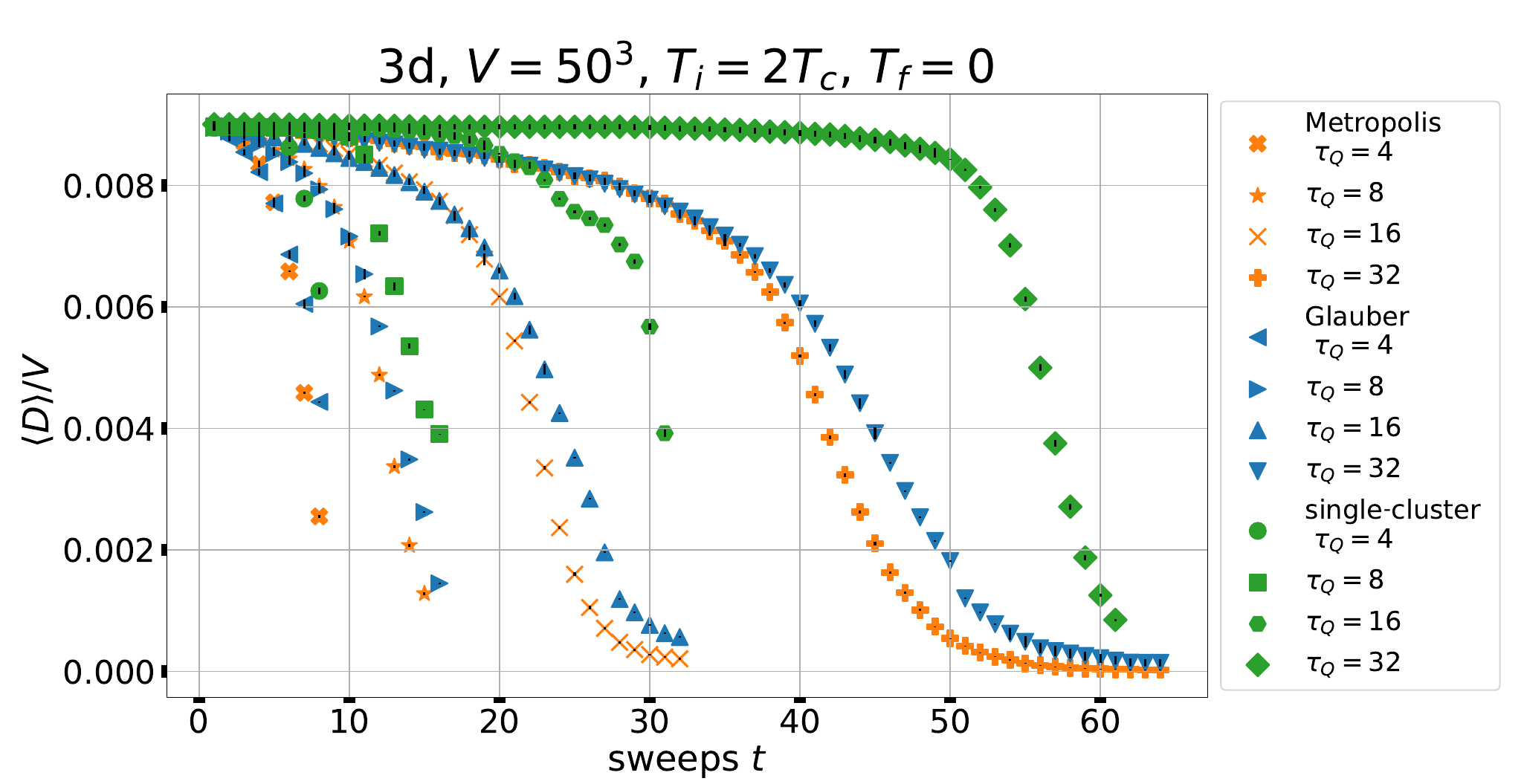}
\caption{\small Markov time evolution of the domain number density
$\la D(t) \ra/V$ during the cooling processes in $d=1$ (top),
$d=2$ (center) and $d=3$ (bottom).}
\label{Ddense}
\end{center}
\end{figure}

In Figure \ref{Ddense}, we consider the mean domain number density,
{\it i.e.}\ the number of domains divided by the volume, $\la D(t) \ra /V$,
which represents the inverse mean domain size.
Again, we show results for the three dimensions and three algorithms
under consideration, with the expected trend.

\begin{figure}[h!]
\begin{center}
\includegraphics[scale=0.3]{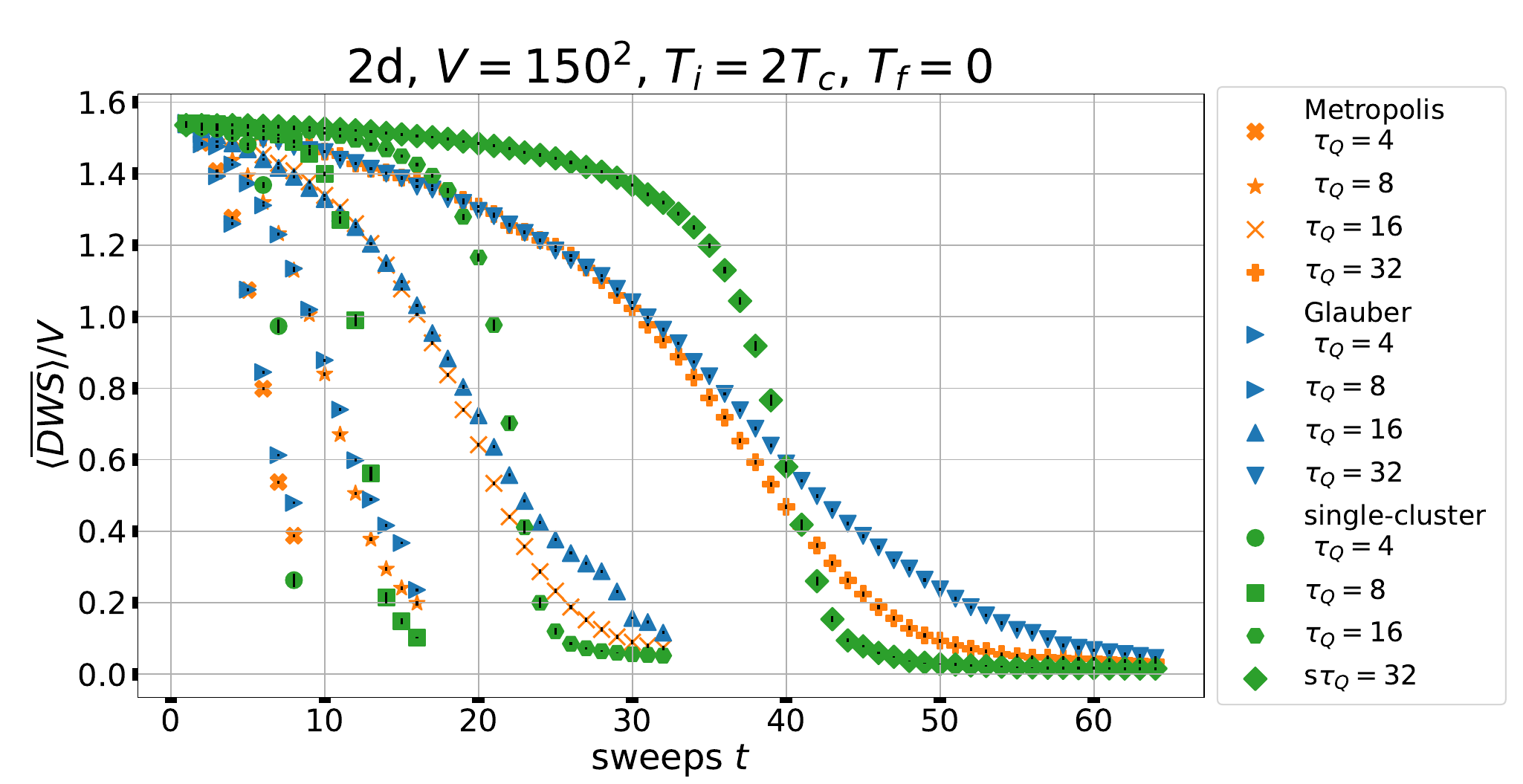}
\includegraphics[scale=0.3]{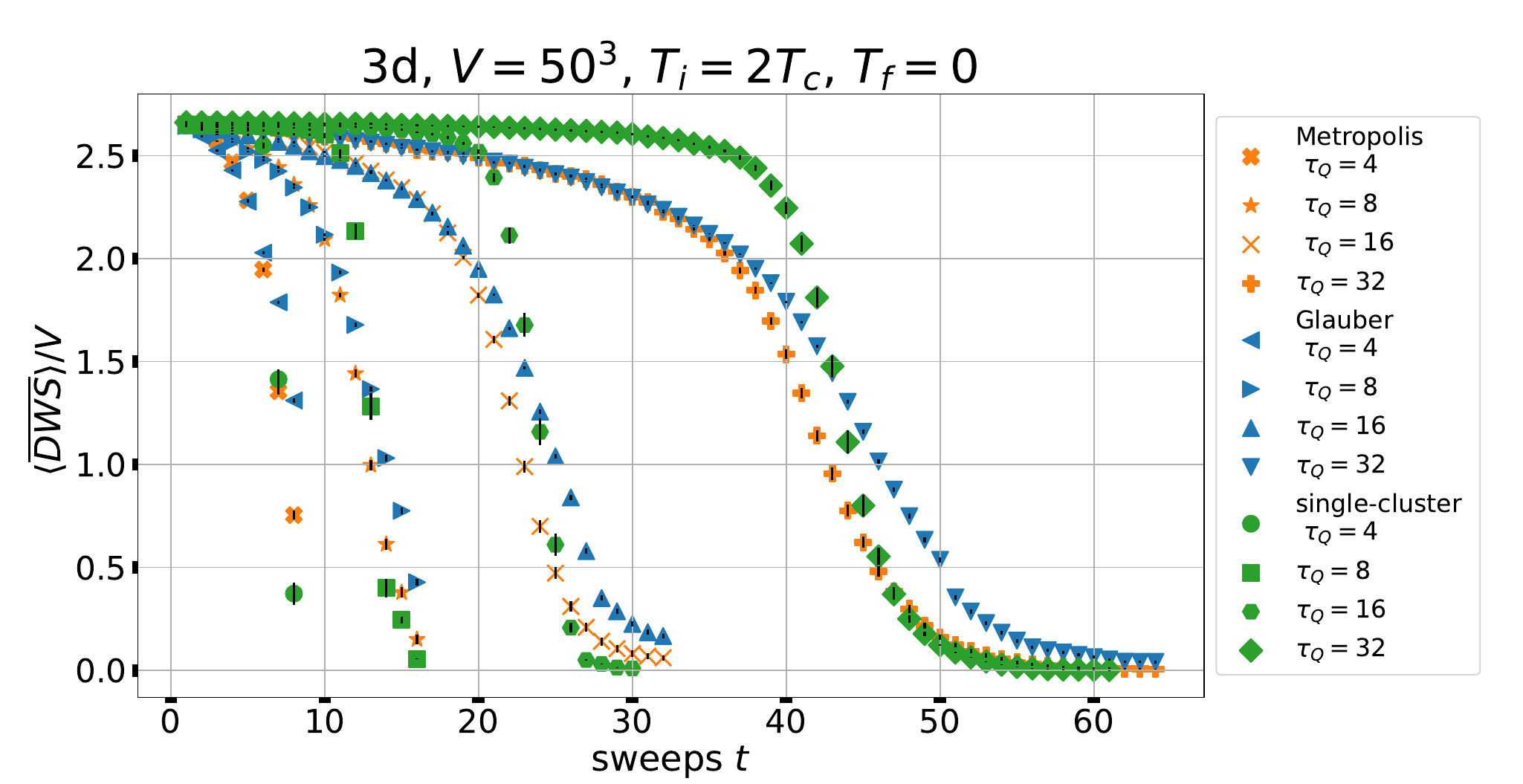}
\caption{\small Markov time evolution of the domain wall size density
$\la \overline{DWS}(t) \ra/V$ during the cooling processes in dimension
$d=2$ (above) and $d=3$ (below).}
\label{DWSdense}
\end{center}
\end{figure}

We proceed to the average domain wall size density,
$\la \overline{DWS} (t) \ra/V$, which is a sensible quantity
only in $d>1$ (cf.\ Section 2). Its Markov time evolution
under cooling is shown in Figure \ref{DWSdense}. It could theoretically
drop down to 0 (equilibrium value at $T=0$), and we see that slow
cooling does take it close to this minimum, in contrast to fast cooling.

Having illustrated the evolution of several observables under cooling
out of thermal equilibrium, we now focus on their values when the
system arrives at its final temperature $T_{\rm f}=0$. Based on Zurek's
conjecture, we are particularly interested in possible scaling-laws,
which relate these remnant quantities to some (negative)
power of the inverse cooling rate $\tau_{Q}$,
\be  \label{scalaw}
\left\{ \frac{1}{V} \Big\la D(t = \tau_{Q}) \Big\ra \quad {\rm or} \quad
\frac{1}{V} \Big\la \overline{DWS} (t = \tau_{Q}) \Big\ra \right\}
\propto \tau_{Q}^{-b} \ .
\ee

\begin{figure}[h!]
\begin{center}
\includegraphics[scale=0.31]{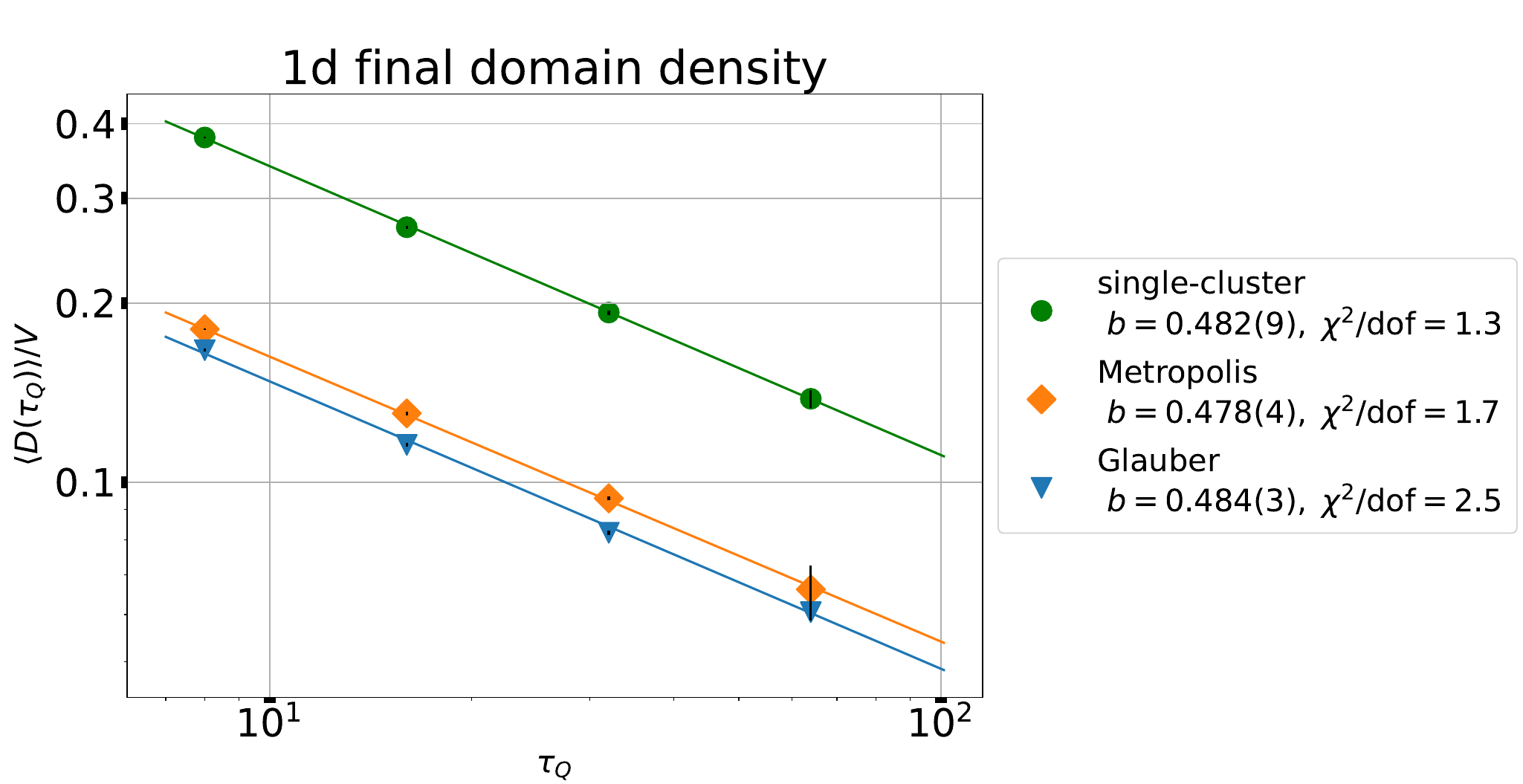}
\includegraphics[scale=0.31]{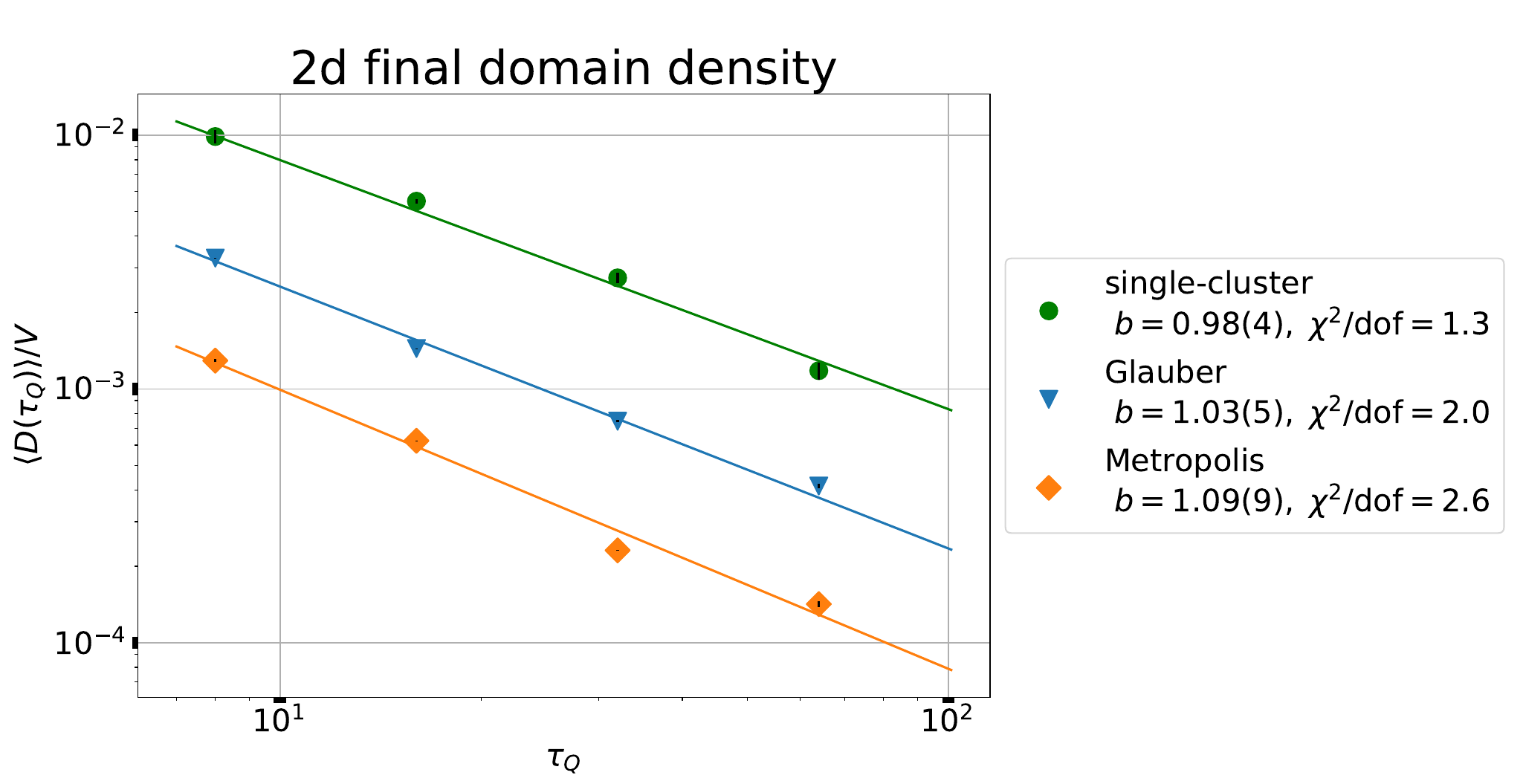}
\includegraphics[scale=0.31]{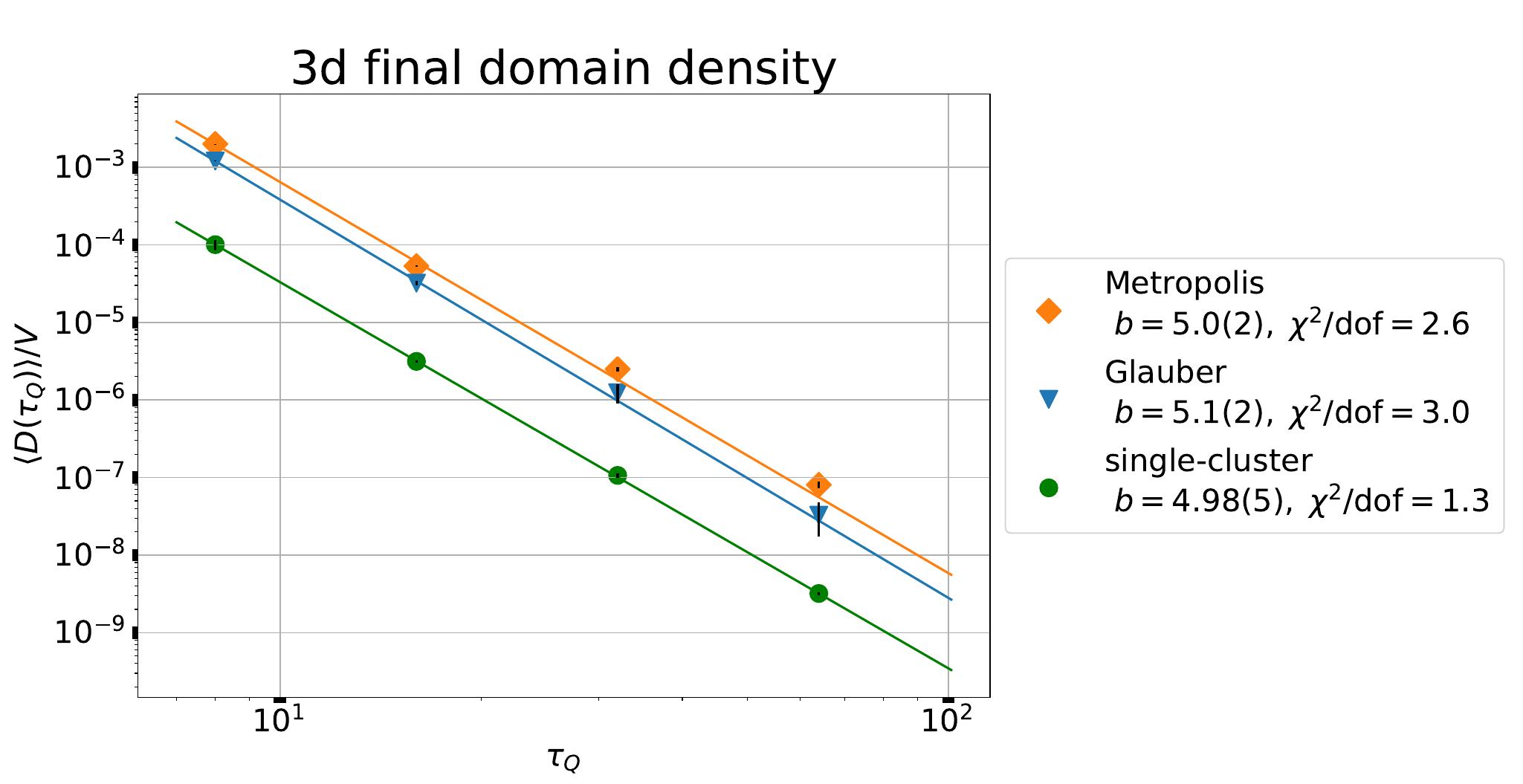}
\caption{\small The domain number density at the end of the cooling
processes (at $t=\tau_{Q}$), $\la D(\tau_{Q})\ra /V$, in $d=1,2$ and
$3$, for the Metropolis algorithm, the Glauber algorithm and
the single-cluster algorithm.}
\label{Dfin}
\end{center}
\vspace*{-8mm}
\end{figure}

Figure \ref{Dfin} shows such scaling-law fits for the final domain
number density $\la D(\tau_{Q}) \ra /V$. In dimensions $d=1, 2$ and $3$,
and for all three algorithms, we see that the data
are well compatible with scaling-laws. The exponents are indicated at
the right-hand side of the plots (the uncertainty of the last digit
is added in parentheses). We also indicate in each case the ratio $\chi^{2}$
by the number of degrees of freedom (dof), as the usual measure of
the fitting quality. This ratio is $\leq 3.0$ in all cases,
so all these fits are fully satisfactory.

\begin{figure}[!h]
\begin{center}
\includegraphics[scale=0.31]{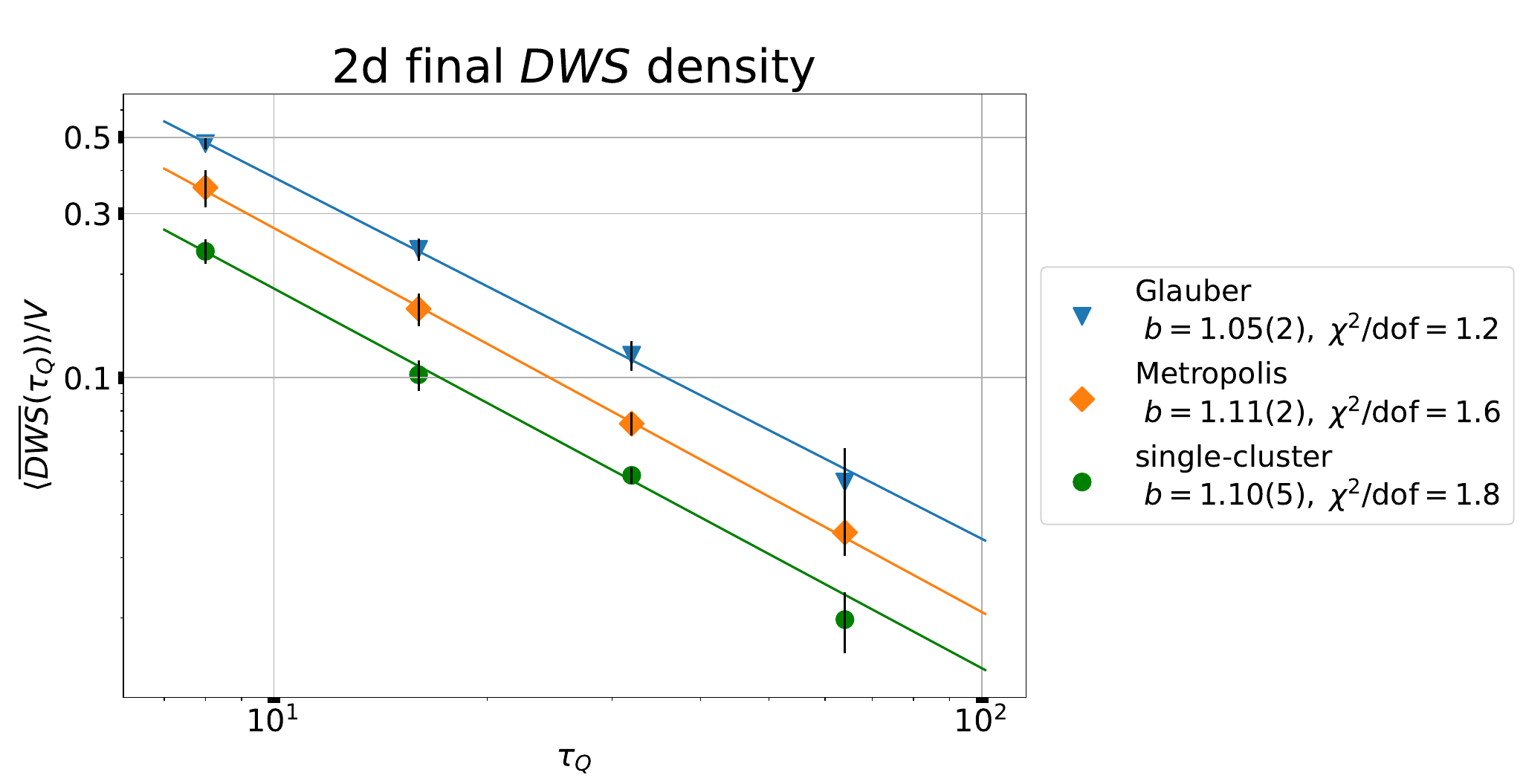}
\includegraphics[scale=0.31]{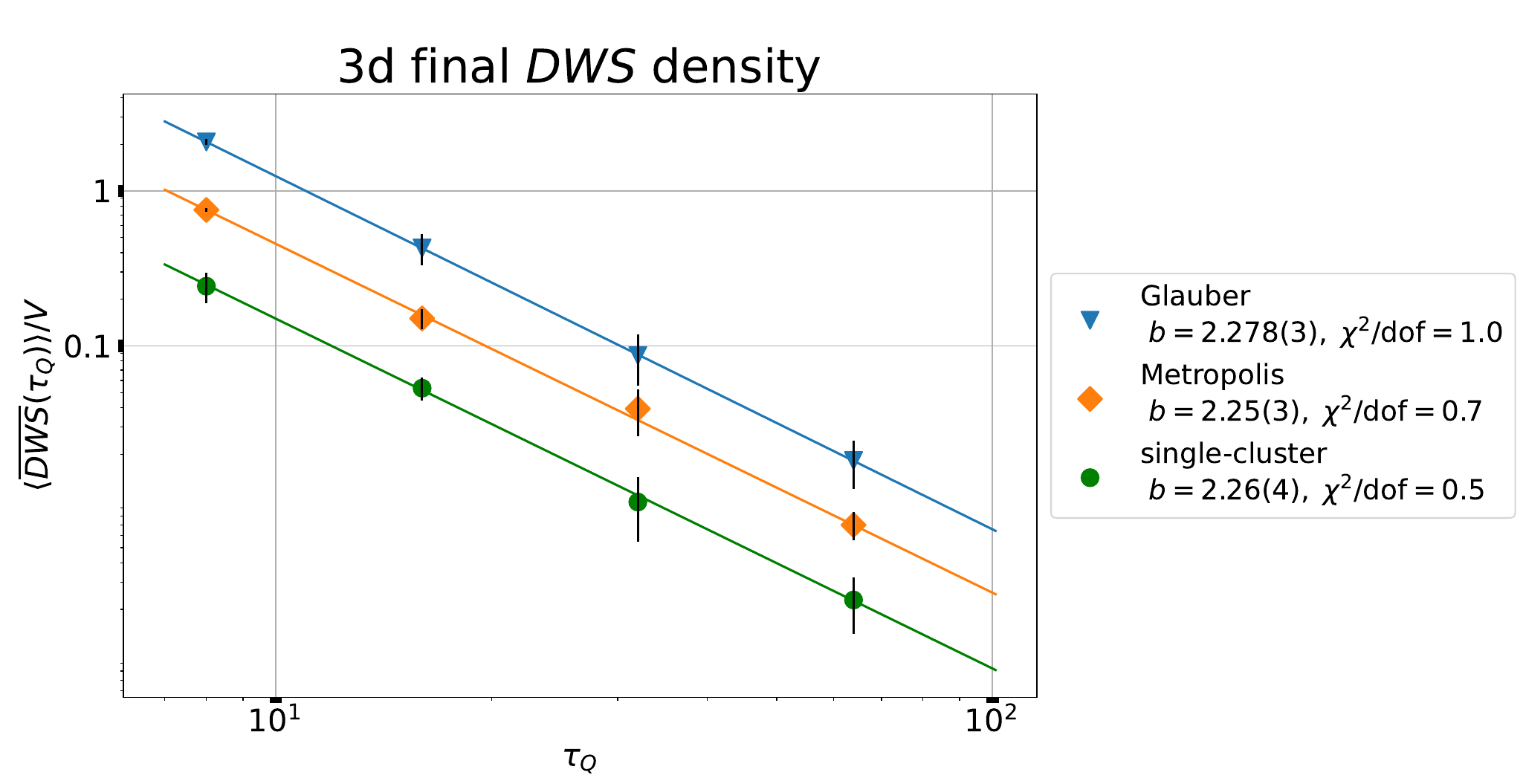}
\caption{\small The domain wall size density at the end of the
cooling processes, $\la \overline{DWS} (\tau_{Q}) \ra /V$, in $d=2$
and $3$, for the Glauber algorithm, the Metropolis algorithm and
the single-cluster algorithm.}
\label{DWSfin}
\end{center}
\vspace*{-6mm}
\end{figure}

Finally, Figure \ref{DWSfin} illustrates the scaling-law for
$\la \overline{DWS}(t=\tau_{Q}) \ra /V$ in $d=2$ and $3$
(using the same notation as in Figure \ref{Dfin}). Again,
the proportionality ansatz (\ref{scalaw}) works well in both
dimensions and for all three algorithms.

The exponents obtained by the scaling-law fits in Figures \ref{Dfin}
and \ref{DWSfin} are summarized in Table \ref{tabexp}. These exponents
do not depend on the exact choice of the initial temperature $T_{\rm i}$.
Moreover, we see a striking consistency between the three algorithms,
which is highly remarkable, especially because the single-cluster
algorithm works in a manner, which is very different from the local
update algorithms.\footnote{The local update algorithms are assumed
  to be closest to a physical cooling process, which is due to microscopic
  dynamics \cite{Lin14}.}

The value for $b$ that we obtained in $d=1$ for $\la D \ra /V$ is close to
$1/2$. In fact, that exponent can be estimated based on the relation
$D \propto \tau_{Q}^{-1/z}$, where $z$ is the dynamical critical exponent,
and random walk dynamics of the 1d domain walls, or kinks,
suggests $z \simeq 2$ \cite{Mayo21}.
The exponent $b \simeq 1/2$ was also measured in the Josephson junction
experiment \cite{Monaco}. That system can be considered
as (essentially) 1d too, but other properties are very different
({\it e.g.}\ the superconducting phase breaks a local, continuous U(1)
symmetry). This exponent disagrees with the straight theoretical
prediction of 1/4, but the authors of Ref.\ \cite{Monaco} manage to
modify the theoretical derivation such that they arrive at 1/2.

\begin{table}[h!]
\begin{center}  
\begin{tabular}{|c|c||c|c|c|}
\hline
observable & dimension & \multicolumn{3} {c}{algorithm} \\
\cline{3-5}
 & & Glauber & Metropolis & single-cluster \\
\hline
\multirow{3}{*}{$\la D \ra /V$} & 1d & 0.482(9) & 0.478(4) & 0.484(3) \\
               & 2d & 1.03(5) & 1.09(9) & 0.98(4) \\
               & 3d & 5.1(2)  & 5.0(2)  & 4.98(5) \\
\hline
\multirow{2}{*}{$\la \overline{DWS} \ra /V$}
& 2d & 1.05(2) & 1.11(2) & 1.10(5) \\
                            & 3d & 2.278(3)& 2.25(3) & 2.26(4) \\
\hline
\end{tabular}
\end{center}
\vspace*{-2mm}
\caption{Values of the exponent $b$ in the scaling-laws (\ref{scalaw})
for the final values of two types of domain-related densities, at the
end of a rapid cooling process down to $T_{\rm f}=0$
(at $t=\tau_{Q}$, cf.\ eq.\ (\ref{Ttlin})).
\label{tabexp}}
\end{table}

\section{Conclusions}

We have performed Monte Carlo simulations of the Ising model ---
with classical spin variables ---
out of thermal equilibrium, in order to test Zurek's scaling-law
conjecture in an unusual and unexpected setting. In dimensions $d=1,2$
and $3$, and for three different algorithms, it is accurately
confirmed for two domain-related observables.
In this regard, our 3d simulations are in line with the experimental
results of Ref.\ \cite{Du23}. It is natural, however, that the
experimentally measured exponent $b$ for domain number density differs
from our value, since it refers to a finite temperature $T_{\rm f}$.

Our results for the exponents in the scaling-laws are in good
agreement for the different algorithms.
This suggests that they might be physically meaningful in
Statistical Mechanics, for instance regarding the relaxation time
of systems in the Ising universality class.

We cannot relate these exponents to the predicted formula
\cite{DelCampo1314}.
This prediction refers to a final temperature $T_{\rm f} > 0$,
which matches a certain criterion (end of the quasi-frozen
temperature interval around $T_{\rm c}$), which is, however, hard to
single out. A detailed study in the 3d XY model explored this criterion
based on the autocorrelation time, but it arrived at exponents clearly
below the prediction \cite{Edgar}.
The exact value of $T_{\rm f}$ is crucial in this regard, because the
exponent $b$ varies over a broad range, depending on $T_{\rm f}$.

Still, it is an interesting conclusion that such scaling-laws hold
more generally than expected, as we showed in the Ising model
(which does not have topological defects), and even in its 1d version
(which does not even undergo a phase transition). This encourages numerical
studies of this property in more sophisticated models, with topological
defects and even with topological charges; such Monte Carlo
simulations are in progress.

\ \\
\noindent
{\bf Acknowledgments:}
We are indebted to Jos\'{e} Antonio Garc\'{\i}a-Her\-n\'{a}n\-dez,
Edgar L\'{o}pez-Contreras, Victor Mu\~{n}oz-Vitelly, Jaime Fabi\'{a}n
Nieto Castellanos and El\'{\i}as Natanael Polanco-Eu\'{a}n
for inspiring discussions and advice in the course of the realization
of this project.
The simulations were performed on the cluster of the Instituto
de Ciencias Nucleares, UNAM.

\end{document}